\documentclass[prb,twocolumn,superscriptaddress,showpacs,preprintnumbers,amsmath,amssymb]{revtex4}
\usepackage{mathbbol}
\usepackage{graphicx} 
\usepackage{dcolumn}
\usepackage{bm}

\begin{document}
\bibliographystyle{plain}
\newcommand*{\cm}{cm$^{-1}\ $\,}
\newcommand*{\Tc}{T$_c$\,}
\newcommand*{\sample}{$\mathrm{Ba_2Ti_2Fe_2As_4O}\ $\,}
\newcommand*{\R}{${R(\omega)}\ $\,}
\title{Coexistence of superconductivity and density wave in $\mathrm{Ba_2Ti_2Fe_2As_4O}$: an optical spectroscopy study}

\author{H. P. Wang}
\affiliation{Beijing National Laboratory for Condensed Matter Physics, Institute of Physics, Chinese Academy of Sciences, Beijing 100190, China}

\author{Y. L. Sun}
\affiliation{Department of Physics, Zhejiang University, Hangzhou 310027, China}

\author{X. B. Wang}
\author{Y. Huang}
\author{T. Dong}
\affiliation{Beijing National Laboratory for Condensed Matter Physics, Institute of Physics, Chinese Academy of Sciences, Beijing 100190, China}

\author{R. Y. Chen}
\affiliation{International Center for Quantum Materials, School of Physics, Peking University, Beijing 100871, China}

\author{G. H. Cao}
\affiliation{Department of Physics, Zhejiang University, Hangzhou 310027, China}

\author{N. L. Wang}
\email{nlwang@pku.edu.cn} \affiliation{International Center for Quantum Materials, School of Physics, Peking University, Beijing 100871, China}
\affiliation{Collaborative Innovation Center of Quantum Matter, Beijing, China}


\begin{abstract}
We performed an optical spectroscopy measurement on single crystals of $\mathrm{Ba_2Ti_2Fe_2As_4O}$, which is a newly discovered superconductor
showing a coexistence of superconductivity and density wave orders. The study reveals spectral changes associated with both density wave and superconductivity phase transitions. The density wave phase transition at $T_{DW}\approx$125 K leads to the reconstruction of Fermi surfaces which removes about half of Drude spectral weight. The ratio of 2$\Delta_{DW}$/$k_B
T_{DW}\approx$ 11.9 is considerably larger than the mean-field value based on the weak-coupling BCS theory. At the lowest temperature in the
superconducting state, further spectral change associated with the superconducting condensate is observed. The low frequency optical conductivity could be well modeled within Mattis-Bardeen approach with two isotropic gaps of $\Delta_1(0) =3.4$ meV and $\Delta_2(0)=7.9$ meV. The superconducting properties of $\mathrm{Ba_2Ti_2Fe_2As_4O}$ compound are similar to those of $\mathrm{BaFe_{1.85}Co_{0.15}As_2}$.
\end{abstract}

 \maketitle
\section{Introduction}

The discovery of iron-based superconductors\cite{Hosono} has generated tremendous interest in condensed-matter community and triggered new efforts in
exploring superconductors in new families of compounds. The exploration has been mainly focused on materials with structural, electronic, and
magnetic properties similar to those of the iron or cuprate superconductors, or those close to charge- or magnetic instabilities. The efforts lead to
the discovery of superconductivity in BaTi$_2$Sb$_2$O with $T_c \approx$ 1.2 K \cite{Yajima}. The compound belongs to a two-dimensional (2D) titanium
oxypnictide family, consisting of alternate stacking of conducting octahedral layers Ti$_2$Pn$_2$O (Pn=As, Sb) and other insulating layers (e.g.
Na$_2$, Ba, (SrF)$_2$, (SmO)$_2$) \cite{Adam,Axtell,Ozawa1,Ozawa2,Ozawa3,Ozawa4,Liu1,Wang,Liu2,Zhai}. The undoped compounds in this family commonly
exhibit phase transitions below certain temperatures (e.g. 320 K for Na$_2$Ti$_2$As$_2$O \cite{Axtell,Ozawa1,Ozawa2,Ozawa3,Liu1}, 114 K for
Na$_2$Ti$_2$Sb$_2$O \cite{Ozawa1}, 200 K for BaTi$_2$As$_2$O \cite{Wang}, 45 K for BaTi$_2$Sb$_2$O \cite{Yajima}), as characterized by the sharp
jumps in resistivity and drops in magnetic susceptibility. First principle band structure calculations indicate that the phase transitions are driven
by density wave (DW) instabilities arising from the nested electron and hole Fermi surfaces (FSs) \cite{Pickett,Biani,Singh,Yan,Subedi}. However,
there is no agreement on whether the DW is a charge density wave (CDW) or a spin density wave (SDW). As superconductivity emerges only in compound
with low phase transition temperature, \emph{i.e.} BaTi$_2$Sb$_2$O, and T$_c$ is further enhanced when the phase transition temperature was
suppressed by doping, e. g. Na ( $T_c \approx$5.5 K)\cite{Doan}, K ($T_c \approx$6.1 K) \cite{Pachmayr}, Rb ($T_c \approx$5.4 K)\cite{Rohr} and Sn (
$T_c \approx$2.5 K)\cite{Nakano} and with isovalent Bi substitution for Sb ($T_c \approx$4.6 K)\cite{Zhai}, the family offers a new playground to
study the interplay between superconductivity and DW instabilities.

Among the newly discovered oxypnictide superconductors, \sample (Ba22241) is particularly interesting. From the structural point of view, the
compound can be considered as an intergrowth of $\mathrm{BaFe_2As_2}$ and $\mathrm{BaTi_2As_2O}$ \cite{Cao1}.
Although neither $\mathrm{BaFe_2As_2}$ nor $\mathrm{BaTi_2As_2O}$
is superconducting, the combined structure, Ba22241, shows superconductivity at $T_c$=21.5 K without doping. Furthermore, the compound still shows a
DW transition at $T_{DW}$=125 K. Compared with BaTi$_2$As$_2$O, the DW ordering temperature is significantly reduced. Previous studies show that the
anomaly at $T_{DW}$ is due to a SDW or CDW transition in the $\mathrm{Ti_2As_2O}$ block and the superconductivity arises from
$\mathrm{Fe_2As_2}$ layer rather than $\mathrm{Ti_2As_2O}$ layer\cite{Cao1,Jiang}. First principle calculations indicate the FSs can be separated
into Ti-related sheets and Fe-related sheets and there is a $0.12e$ charge transfer from $\mathrm{Ti_2As_2O}$ layers to $\mathrm{Fe_2As_2}$
layers\cite{Jiang}. Therefore, there is a self-doping effect, which suppresses the stripe type antiferromagnetism at the Fe sites and simultaneously
induces superconductivity. The Ba22241 material represents a rare example to show two-dimensional DW ordering sandwiched by superconducting layers.
It would be very interesting to investigate the charge dynamical properties of this material.

In this work, we report an optical study on \sample single crystals. Above $T_{DW}$, we observe a metallic response with plasma frequency of 2.6 eV.
A typical density wave energy gap forms when the compound enters into the DW ordering phase, leading to the removal of parts of the FSs. The sample
remains metallic below $T_{DW}$. A ratio of $2\Delta_{DW} /k_BT_c \approx$ 11.9 is obtained. The value is considerably larger than the mean-field value of
the weak-coupling BCS theory. Below $T_c$, we observe the formation of superconducting condensate. About $35\%$ of the free carrier
spectral weight in the normal state collapses into the superconducting condensate, suggesting that the material is not in the clean limit. The optical conductivity at the lowest
temperature could be well reproduced by Mattis-Bardeen formula with two isotropic gaps of $\Delta_1(0) =3.4$ meV and $\Delta_2(0)=7.9$ meV. The coexistence of
density wave and superconductivity makes \sample a promising candidate to study collective excitations in broken symmetry states.

\section{Experiments}
The \sample single crystal samples used in our optical measurements were grown by employing a $\mathrm{Ba_2As_3}$ flux method. Detailed procedure of
crystal growth and characterization could be found elsewhere \cite{Cao2}.  The optical reflectance measurements were performed on Bruker IFS 113v and
80v spectrometers in the frequency range from 30 to 35 000 \cm. An \emph{in situ} gold and aluminium overcoating technique was used to obtain the
reflectivity R($\omega$). The real part of conductivity $\sigma_1(\omega)$ is obtained by the Kramers-Kronig transformation of R($\omega$). The
Hagen-Rubens relation was used for low frequency extrapolation; at high frequency side a $\omega^{-1}$ relation was used up to 300 000 \cm, above
which $\omega^{-4}$ was applied.

\section{Results}

\subsection{Density wave state}

\begin{figure}
\includegraphics[height=8.5cm,angle=270]{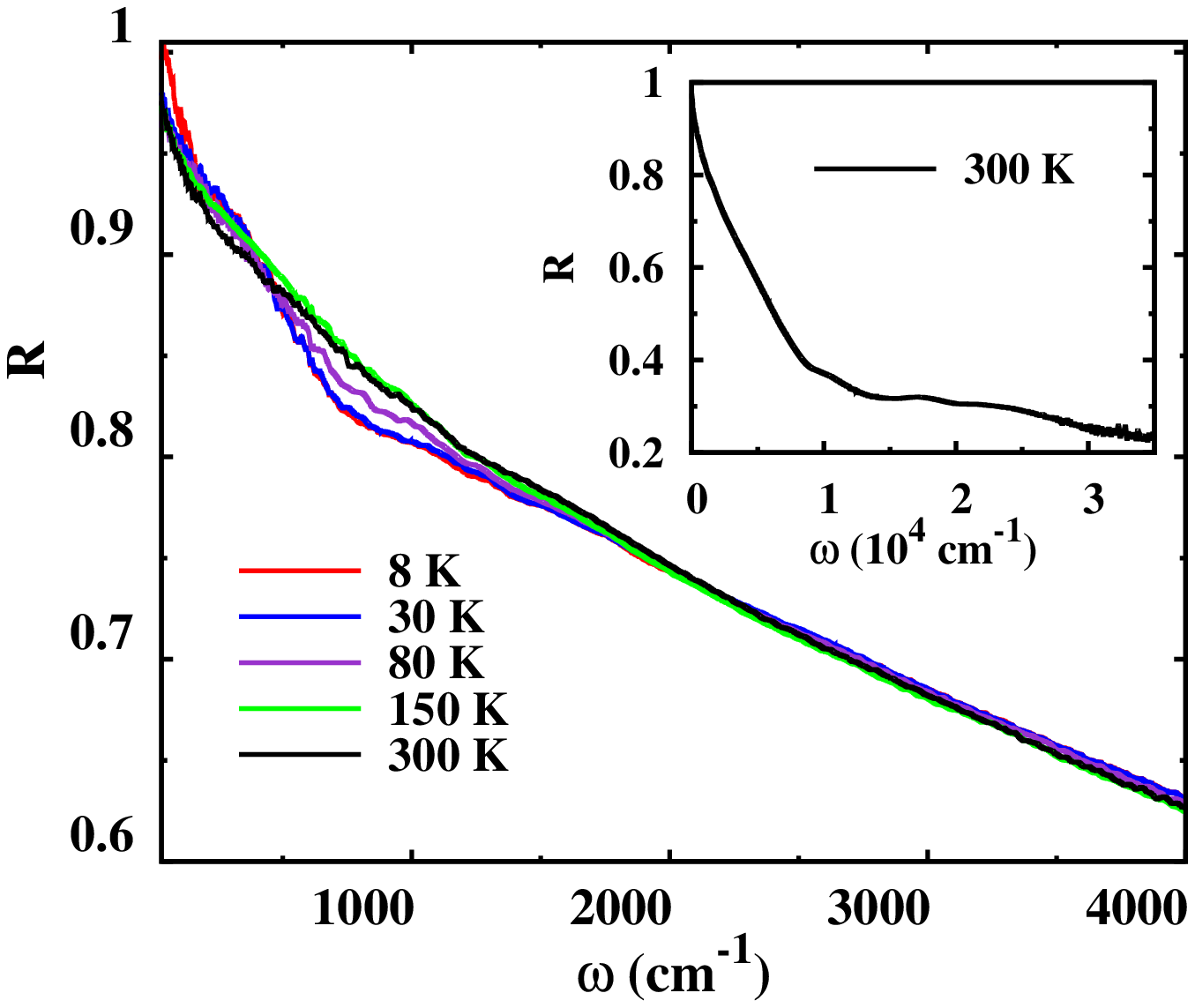}
\caption{(Color online) Optical reflectivity \R between the frequency range 30 \cm and 3000 \cm at five representative temperature. The inset:
${R(\omega)}$ at 300K up to 35000 \cm in a linear frequency range. \label{fig1}}
\end{figure}

\begin{figure}
\includegraphics[height=8.5cm,angle=270 ]{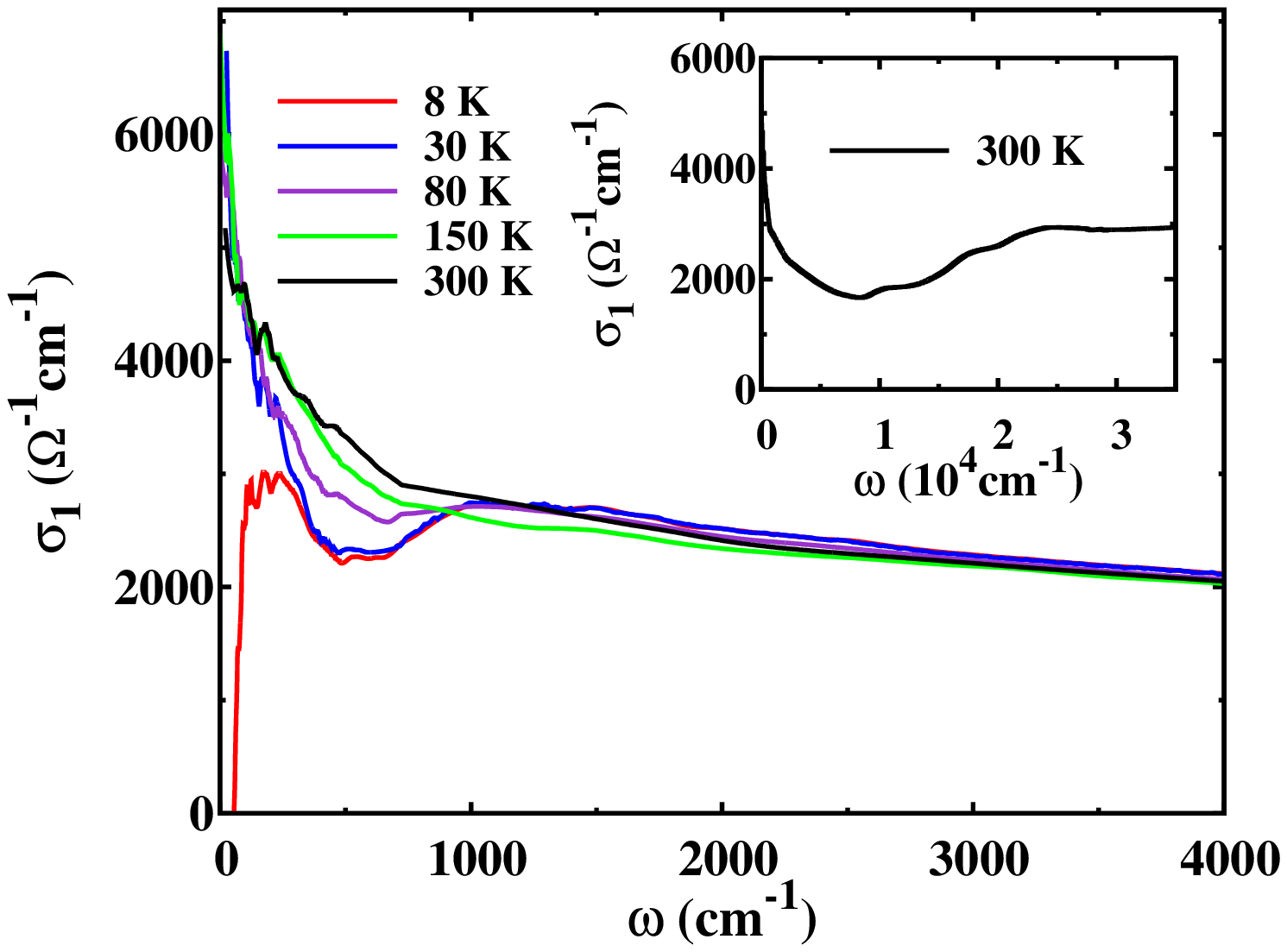}
\caption{(Color online)The real part of optical conductivity $\sigma_1(\omega)$ for \sample below 3000 \cm. The inset: $\sigma_1(\omega)$ up to
35000\cm for 300K. \label{fig2}}
\end{figure}

Figure \ref{fig1} and \ref{fig2} show the reflectivity \R  and real part of the optical conductivity $\mathrm{\sigma_1 (\omega)}$ spectra at
different temperature, respectively. At 300 K and 150 K, the temperature higher than the phase transition, \R shows typical metallic responses. The
\R decreases almost linearly with frequency at low energy region, which is a common feature in materials with strong carrier scatterings. Below the
density wave transition temperature $T_{DW}$, \R is obviously suppressed at the energy roughly lower than 1100 \cm. Accordingly, $\mathrm{\sigma_1 (\omega)}$
shows a peak structure at this frequency. This feature is similar to what has been observed for the two other titanium oxypnictide compounds
Na$_2$Ti$_2$Sb$_2$O \cite{Yue-NTSO} and Na$_2$Ti$_2$As$_2$O \cite{Yue-NTAO}, though its structure is less prominent. The optical data clearly
indicate the formation of a density wave energy gap. At lower energies, \R becomes even higher than the values at higher temperatures and there
still exists well defined Drude component in $\mathrm{\sigma_1 (\omega)}$ .
Therefore, the compound is still metallic and the Fermi surfaces are only partially gapped below $T_{DW}$. The suppressed spectral weight due to the
opening of the density wave energy gap is transferred to higher energy scales.

In Fig. 3, we plot the  frequency-dependent integrated spectral weight at different temperatures in the normal state. The spectral weight is defined
as $W_s=\int_0^{\omega_c} \sigma_1(\omega) d\omega$, where $\omega_c$ is a cut-off frequency. We can see that the spectral weight is gradually
recovered at high frequencies. The spectral weight transfer is seen more clearly in the plot of the ratio of the spectral weight at two different
temperatures below and above the density wave transition, e.g. $W_s(30K)/W_s(300K)$, as shown in the inset of Fig. 3. The value of the ratio is
higher than unity at very low frequency, which is apparently due to the higher conductivity values of the narrow Drude component in the density wave
state. The ratio becomes less than the unity at higher energy due to the opening of density wave energy gap in $\sigma_1(\omega)$ spectrum at 30 K.
Eventually, the spectral weight is recovered and the ratio approaches unity at higher energies.

\begin{figure}
\includegraphics[height=8.5cm,angle=270 ]{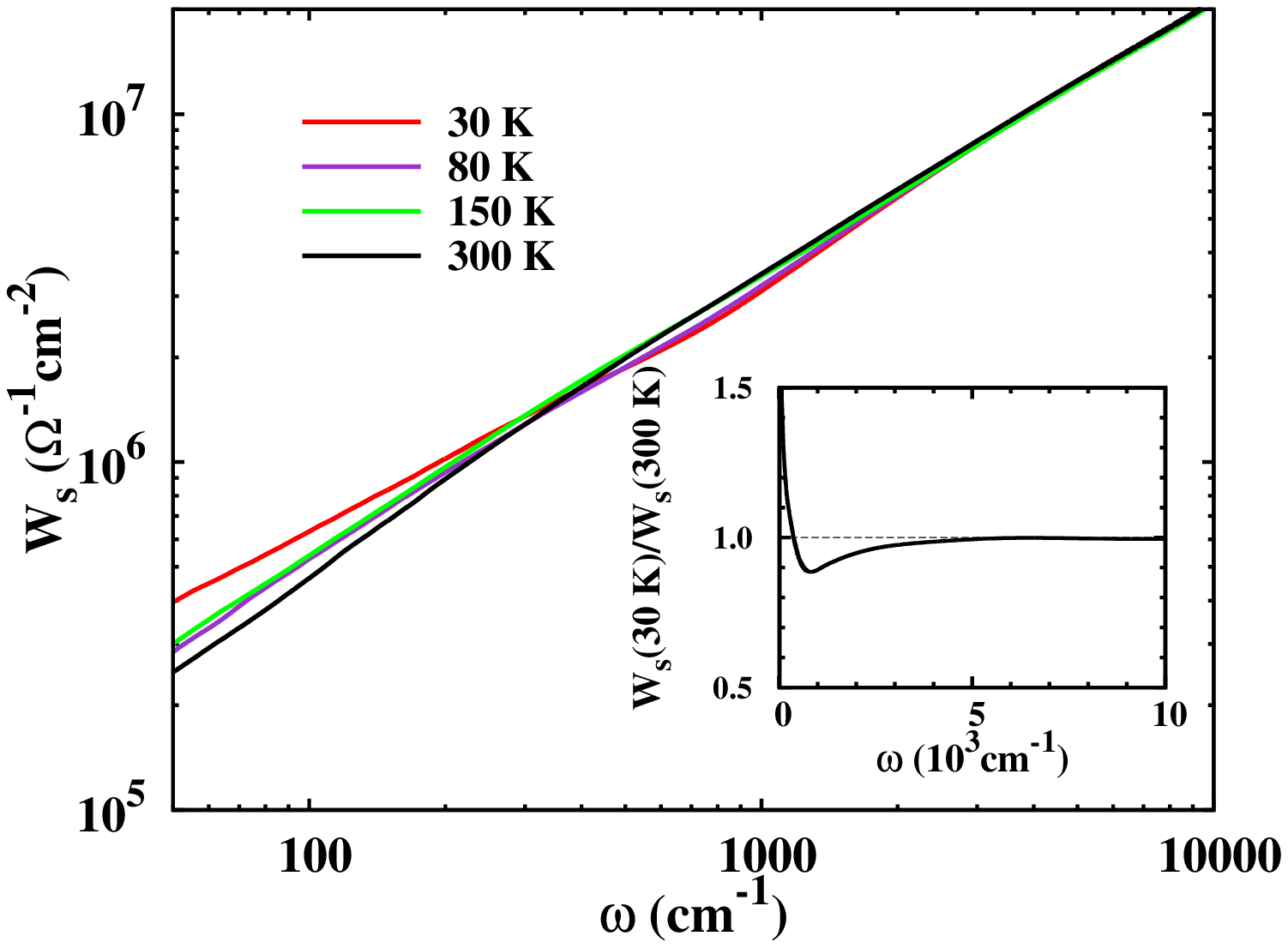}
\caption{(Color online) Cut-off frequency dependent spectral weight at four different temperature. Inset: the normalized spectral weight
W$_s$(30K)/W$_s$(300K) up to 10000 \cm. \label{fig4}}
\end{figure}

\begin{figure}
\includegraphics[height=8.5cm,angle=270 ]{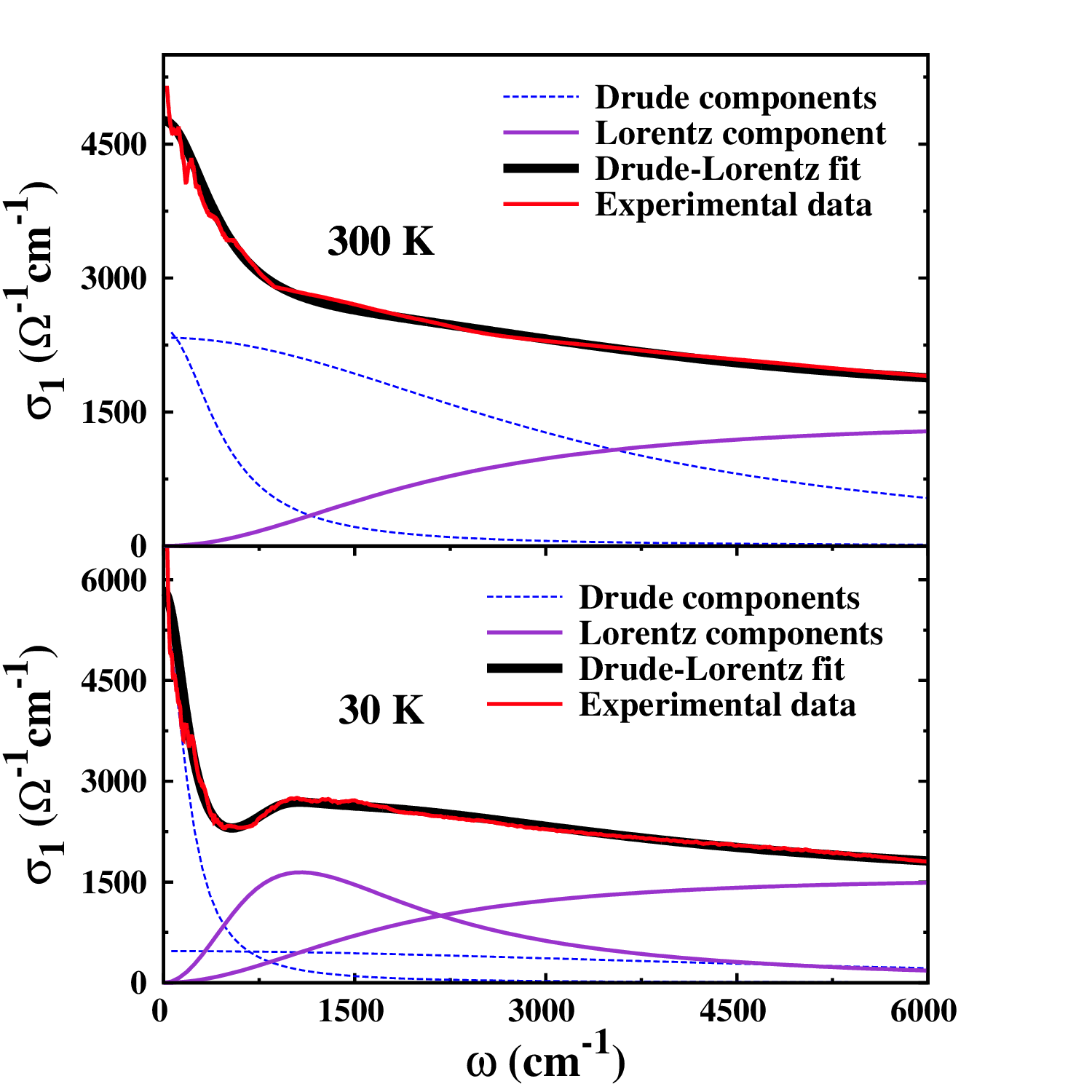}
\caption{(Color online) The experimental data of $\sigma_1(\omega)$ at 300K and 30 K together with the Drude-Lorentz fits. \label{fig3}}
\end{figure}

To characterize the spectral change across the phase transition, we decompose the optical conductivity spectra into different components using a
Drude-Lorentz analysis. The dielectric function has the form \cite{HWZ}
\begin{equation}
    \epsilon(\omega)=\epsilon_{\infty}-\sum\limits_{i} \frac{\omega_{p,i}^2}{\omega^2+i\omega/\tau_i} +\sum\limits_{j} \frac{\Omega_{j}^2}{\omega_{j}^2-\omega^2-i\omega/\tau_j}
\end{equation}
where $\epsilon_\infty $ is the dielectric constant at high energy, the middle and last terms are the Drude and Lorentz components. The complex
conductivity is $\sigma(\omega)=\sigma_1(\omega)+i\sigma_2(\omega)=-i\omega[\epsilon(\omega)-\epsilon_\infty]/4\pi$. As shown in Fig. \ref{fig3}, the conductivity spectrum below 6000 \cm at room temperature could be well reproduced by two Drude components (a sharp one and a broaden one) and a Lorentz component. At 30 K an additional Lorentz
component at low frequency is added to character the density wave energy gap. The two Drude components model has been widely used in the study of
Fe-based superconductors or other multi-band systems \cite{Nakajima,WuDan,Homes}. The overall plasma frequency
$\omega_p$ is considered to contribute from two different channels with
$\omega_p={(\omega_{p1}^2+\omega_{p2}^2)}^{1/2}$. In the present case we obtain $\omega_p\approx$ 21600 \cm above $T_{DW}$ and $\omega_p \approx
$ 15000 \cm at 30 K. Another method to estimate the overall plasma frequency is to calculate the low-$\omega$ spectral weight,
$\omega_p^2=8 \int_0^{\omega_c} \sigma(\omega) d\omega $. The cut-off frequency $\omega_c$ is chosen so as to make the integration cover all
contribution from free carriers and exclude contribution from interband transitions. Usually, the integral goes to a frequency where the conductivity
shows a minimum. We expect there is a balance between the Drude component tail and the onset part of interband transition. In view of the complex
band structure of this material there should exist multiple interband transitions even at mid-infrared frequency region. There is no good way to
choose the cut-off frequency. Then, we refrain from calculating the plasma frequency from this method. Assuming that the effective mass of charge
carriers does not change very much, we get the ratio of $\omega_{p,30k}^2/\omega_{p,300k}^2 \approx 48\%$, suggesting that about half of the free
carrier spectral weight remains after the phase transition. Compared with the $\mathrm{Na_2Ti_2As_2O}$\cite{Yue-NTAO} and
$\mathrm{Na_2Ti_2Sb_2O}$\cite{Yue-NTSO} in which about $95\%$ FSs is removed, here in \sample much less fraction of the FSs is removed after the phase
transition. This result can be easily understood considering the fact that the FSs are contributed from both $\mathrm{Ti_2As_2O}$ and
$\mathrm{Fe_2As_2}$ layers and only those from $\mathrm{Ti_2As_2O}$ layer are affected by the density wave phase transition. The density wave gap
size can be identified by the peak position of the Lorentz component and we get $2\Delta_{DW}\approx 1070$ \cm =134 meV. The gap ratio
$2\Delta_{DW}/k_BT_c \approx 11.9$ is larger than the mean-field value based on the weak-coupling BCS theory. Similar values were also seen for two
other titanium-based compounds in optical studies \cite{Yue-NTSO,Yue-NTAO}.

\subsection{Superconducting state}

\begin{figure}
\includegraphics[height=8.5cm,angle=270 ]{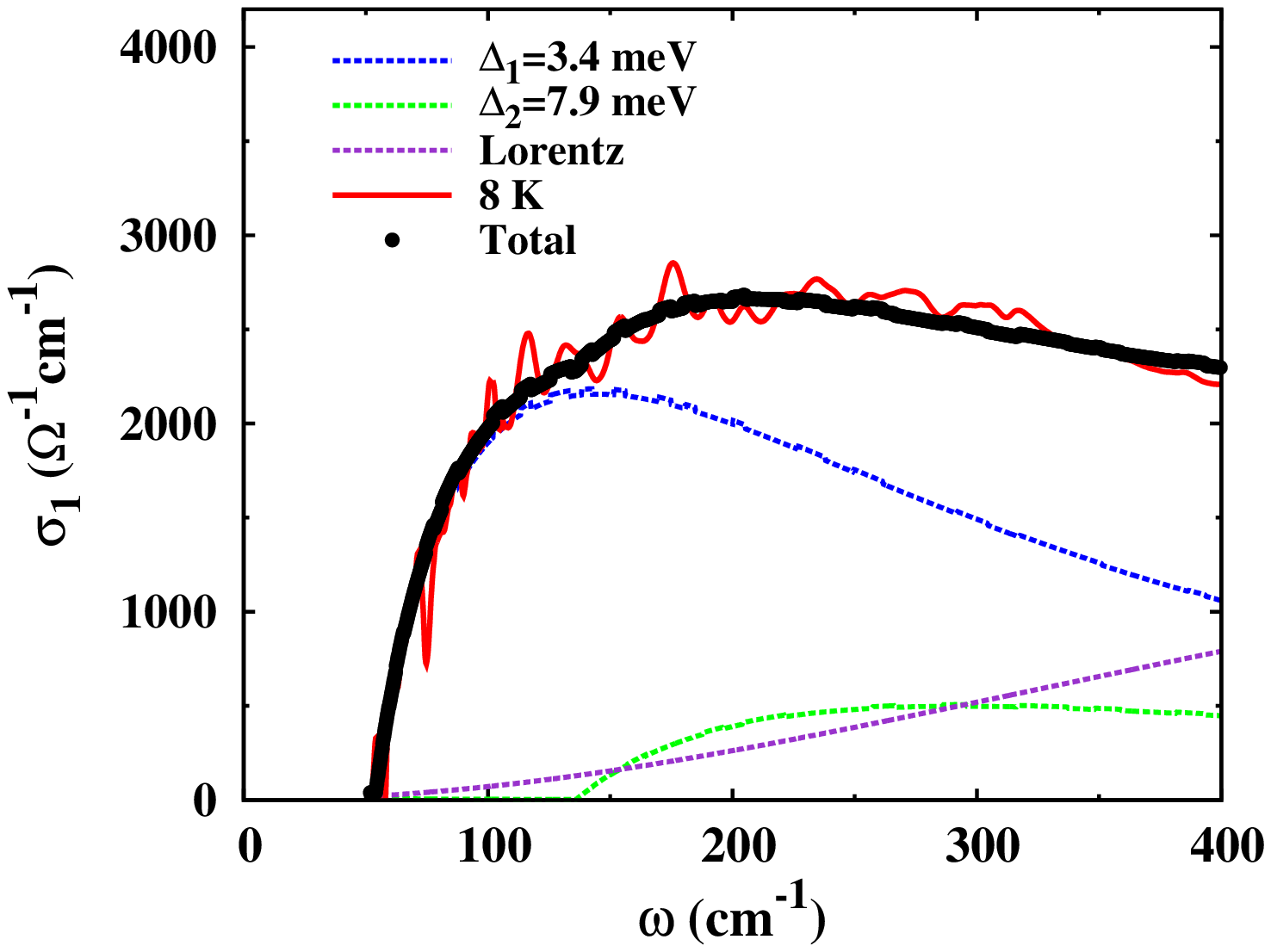}
\caption{(Color online) Up panel: the real part of in-plane optical conductivity of \sample at 8 K together with two gap fits and mid-infrared
componet.\label{fig5}}
\end{figure}

Below $T_c$, further spectral change is observed due to the development of superconductivity. Compared with the spectrum at 30 K in the normal state,
\R at 8 K shows a clear upturn at very low-$\omega$ region and reach approximately unity at about 55 \cm, as can be seen from Fig. 1.
Correspondingly, the $\mathrm{\sigma_1 (\omega)}$ drops to zero at the same energy scale, as shown in Fig. \ref{fig5}. Therefore, superconducting
condensate and formation of superconducting energy gap are clearly observed in far-infrared measurement. According to the Ferrell-Glover-Tinkham
(FGT) sum rule, the difference of the spectral weight between $T\approx T_c$ and $T\ll T_c$ (the so-called missing area) is related to the formation
of a superconducting condensate \cite{Ferrell},
\begin{equation}
W_c\equiv W_s(\omega_c,T\approx T_c)-W_s(\omega_c,T\ll T_c)=\omega^2_{p,S}/8,
\end{equation}
where $\omega_{p,S}$ is the square of the superconducting plasma frequency and $\omega_c$ is the cut off frequency which should be chosen such that
the $\omega^2_{p,S}$ converges smoothly. The equation implies that the spectral weight lost in $\sigma_1(\omega)$ in the superconducting state has
been transferred to the weight at the zero frequency delta function. With $\omega_c\simeq 1000$ \cm, we obtain $\omega_{p,S}$=5350 \cm. The
superconducting condensate can also be extract from $\epsilon_1(\omega)=\epsilon_\infty-\omega^2_{p,S}/ \omega^2$ by determining $[-\omega^2
\epsilon_1(\omega)]^{1/2}$ in the $\omega\to 0$ limit. $\epsilon_1$ can be extracted from the complex optical conductivity, then we get
$\omega_{p,S}$=5430 \cm. The results from two methods suggest the superconducting plasma frequency $\omega_{p,S}\simeq$5400 \cm. The penetration
depth is related to the superconducting plasma frequency by $\lambda$=c/$\omega_{p,S}$, then we get the value of $\lambda \approx 2900
\mathrm{\r{A}}$. We note that the value of $(\omega_{p,S}^2/\omega_{p,D}^2 \simeq 0.35)$, that is, less than half of the low-frequency free carrier
spectral weight in the normal state has collapsed into superconducting condensate. It implies that the superconductor is actually in the dirty limit \cite{Homes}. Therefore,
the optical conductivity can be modeled using a Mattis-Bardeen approach\cite{Mattis,Zimmermann}. As shown in Fig. \ref{fig5}, the conductivity spectrum at
the lowest temperature 8 K can be well reproduced by two isotropic energy gaps with $\Delta_1 =3.4$ meV, $1/\tau_1=10\Delta_0$ and $\Delta_2=7.9$
meV, $1/\tau_2=6\Delta_0$ together with a Lorentzian tail. We note that the relation $1/\tau \gtrsim 2\Delta$ agrees well with the above analysis of
dirty limit superconductivity. For the smaller energy gap, the ratio $2\Delta_1/k_B T_c\simeq 3.7 $, which is close to the BCS weak coupling limit.
But for the larger energy gap, the ratio $2\Delta_2/k_B T_c\simeq 8.5 $ is significantly larger than the BCS mean field value. It is worth noting
that those results are similar to those observed for the electron doped Fe-pnictide compound $\mathrm{BaFe_{1.85}Co_{0.15}As_2}$\cite{Homes,Kim,Heumen}, a prototype iron-based
superconductor. The result is not surprising as the superconductivity in \sample arises from the $\mathrm{Fe_2As_2}$ layers. If we take the charge
transfer from $\mathrm{Ti_2As_2O}$ layer to $\mathrm{Fe_2As_2}$ layer as self-doping, the doping level in FeAs layer is actually close to that for
$\mathrm{BaFe_{1.85}Co_{0.15}As_2}$.

\sample represents a very interesting compound exhibiting the coexistence of superconductivity and density wave instability. At present, there is no
agreement on the nature of the density wave instability, that is, whether it is a spin density wave or charge density wave order. Although the first
principle calculations on some of the titanium oxypnictides, e.g. Na$_2$Ti$_2$As$_2$O and Na$_2$Ti$_2$Sb$_2$O\cite{Yan}, suggest an SDW order driven by
the nesting of disconnected FSs, to date no magnetic order has been detected on any compound in this titanium oxypnictide family. Instead, the
available NMR measurement on some of the compounds, e.g. BaTi$_2$Sb$_2$O\cite{NMR}, actually reveals an absence of internal field at the Sb site,
which therefore favors a CDW origin. If the density wave order in \sample is indeed a CDW order, then the material would provide a promising
candidate to study the collective excitations in the broken symmetry states, in particular, the amplitude mode in the superconducting state. Electronic Raman
scattering is a primary tool to probe the collective amplitude mode in a superconductor, however it is limited to a compound showing the coexistence
of superconductivity and CDW, for example, in NbSe$_2$ \cite{Sooryakumar,Sacuto}. The present compound has much higher $T_c$ than NbSe$_2$, it might be easier to probe such collective excitations.

\section{Summary}
To summarize, we have performed an optical spectroscopy study on \sample single crystal. We observed spectral changes associated with both density wave and superconductivity phase transitions. Compared with two other density wave compounds in the titanium
oxypnictide family, $\mathrm{Na_2Ti_2Sb_2O}$ and $\mathrm{Na_2Ti_2As_2O}$, much weaker spectral change has been found across the density wave phase transition at 125 K in $\mathrm{Ba_2Ti_2Fe_2As_4O}$. This could be attributed to the presence of more Fermi surface sheets being contributed from both $\mathrm{Ti_2As_2O}$ to $\mathrm{Fe_2As_2}$ layers, and only those from $\mathrm{Ti_2As_2O}$ layers are affected by the density wave phase transition. With the development of superconductivity below 21.5 K, further spectral change associated with the superconducting condensate was observed. Our analysis indicates that about $35\%$ of free carrier spectral weight has collapsed into the superconducting condensate at $T\ll T_c$, suggesting that the superconductivity in this compound is in the dirty limit. The low frequency optical conductivity could be well modeled within Mattis-Bardeen approach with two isotropic gaps of $\Delta_1(0) =3.4$ meV and $\Delta_2(0)=7.9$ meV. The coexistence of
density wave and superconductivity makes \sample a promising candidate to study collective excitations in broken symmetry states.

\begin{acknowledgements}
This work is supported by the National Science Foundation of China (11120101003, 11327806), and the 973 project of the Ministry of Science and
Technology of China (2011CB921701, 2012CB821403).
\end{acknowledgements}

\end{document}